\documentclass[fleqn]{llncs}    % Specifies the document style.
\usepackage{times}
\usepackage{latexsym}
\usepackage{amssymb}
\usepackage{xspace}
\usepackage{url}
\usepackage{graphicx}
\usepackage{epsfig}
\usepackage{array}
\usepackage{rotating}
\usepackage{a4,a4wide}

\newcommand{\dlv}{{\small DLV}\xspace}
\newcommand{\dlvdb}{{\small DLV}$^{DB}$\xspace}

\newcommand{\hsp }{\noindent \hspace{0.5cm}}

\newcommand{\aspide}{{\em ASPIDE}\xspace}

  {\end{list}}

\begin{document}

\title{Unit Testing in \aspide}

\author{ Onofrio Febbraro$^1$ \and Nicola Leone$^2$ \and Kristian Reale$^2$ \and Francesco Ricca$^2$}
\institute{
$^1$ DLVSystem s.r.l. - P.zza Vermicelli, Polo Tecnologico, 87036 Rende, Italy \\
\email{febbraro@dlvsystem.com} \\
$^2$Dipartimento di Matematica, Universit{\`a} della Calabria, 87036 Rende, Italy\\
\email{\{leone,reale,ricca\}@mat.unical.it}
}

\maketitle

\begin{abstract}
Answer Set Programming (ASP) is a declarative logic programming formalism,
which is employed nowadays in both academic and industrial real-world applications.
Although some tools for supporting the development of ASP programs
have been proposed in the last few years, 
the crucial task of {\em testing} ASP programs received less attention,
and is an Achilles' heel of the available programming environments.

In this paper we present a language for specifying and running {\em unit tests} on ASP programs.
The testing language has been implemented in \aspide, a comprehensive IDE for ASP, 
which supports the entire life-cycle of ASP development with a collection of user-friendly graphical tools 
for program composition, {\em testing}, debugging, profiling, solver execution configuration, and output-handling.

\end{abstract}

%%%%%%%%%%%%%%%%%%%%%%%%%%%%%%%%%%%%%%%%%%%%%%%%%%%%%%%%%%%%%%%
\section{Introduction}\label{sec:introduction}

Answer Set Programming (ASP)~\cite{gelf-lifs-91} is a declarative logic programming formalism
proposed in the area of non-monotonic reasoning. 
The idea of ASP is to represent a given computational
problem by a logic program whose answer sets correspond to solutions,
and then use a solver to find those solutions \cite{lifs-99a}.

The language of ASP~\cite{gelf-lifs-91} supports a number of modeling constructs including 
disjunction in rule heads, nonmonotonic negation~\cite{gelf-lifs-91}, 
(weak and strong) constraints~\cite{bucc-etal-2000a}, aggregate functions~\cite{fabe-etal-2011-aij}, and more.
These features make ASP very expressive~\cite{eite-etal-97f}, and suitable for developing advanced
real-world applications. ASP is employed in several fields, from 
Artificial Intelligence~\cite{gebs-etal-2007-lpnmr-competition,bald-etal-01,bara-gelf-2000,bara-uyan-2001,fran-etal-2001,noge-etal-2001} to Information Integration~\cite{leon-etal-2005}, and  
Knowledge Management~\cite{bara-2002,grass-etal-09-apps-lpnmr}.
Interestingly, these applications of ASP recently have stimulated some interest also in industry~\cite{grass-etal-09-apps-lpnmr}.

On the one hand, the effective application of ASP in real-world scenarios was made possible by 
the availability of efficient ASP systems%
~\cite{gebs-etal-2007-lpnmr-competition,denecher-etal-2009-lpnmr-competition,cali-etal-2010-competition-lpnmr}. 
On the other hand, the adoption of ASP can be further boosted by offering 
effective programming tools capable of supporting the programmers 
in managing large and complex projects~\cite{dovi-erd}.
%The most diffused programming languages are always placed side by side with 
%SDKs featuring a rich set of tools
%Conversely, the lack of effective programming tools may discourage the usage of 
%the ASP programming paradigm~\cite{dovi-erd}, 
%even if it could provide the needed reasoning capabilities
%at a lower (implementation) price than traditional imperative languages.

%Indeed, People interested in programming, in general, wish to have 
%a set of editing features
%(which are already available for a long time in development 
%tools for imperative languages) like syntax coloring, syntax highlighting, code completion, error 
%management, quick fix, etc. Debugging, profiling and testing tools, 
%as well as the capability of organizing programs in projects are
%fundamental for assisting the development of complex applications.

In the last few years, a number of tools for developing ASP programs have been proposed, 
including editors and debuggers%
~\cite{perri-etal-07-sea,Sureshkumar-etal-07,brain-etal-07-sea,brai-devo-2005,elka-etal-2005,oetsch-etal-2010,brai-etal-2007-lpnmr,devo-scha-2007-sea,devo-scha-2009-sea,ricc-etal-2008-jlc,febb-etal-lpnmr11}.
Among them, \aspide~\cite{febb-etal-lpnmr11} 
--which stands for Answer Set Programming Integrated Development Environment--
is one of the most complete development tools% 
\footnote{For an exaustive feature-wise comparison with existing environments for developing
logic programs we refer the reader to~\cite{febb-etal-lpnmr11}.}
and it integrates a cutting-edge editing tool
(featuring dynamic syntax highlighting, on-line syntax correction, 
autocompletion, code-templates, quick-fixes, refactoring, etc.)
with a collection of user-friendly graphical tools for program composition, debugging, 
profiling, DBMS access, solver execution configuration and output-handling.

Although so many tools for developing ASP programs have been proposed up to now, 
the crucial task of {\em testing} ASP programs received less attention~\cite{janh-2010,janh-2011},
and is an Achilles' heel of the available programming environments.
Indeed, the majority of available graphic programming environments for ASP does not provide 
the user with a testing tool (see~\cite{febb-etal-lpnmr11}), 
and also the one present in the first versions of \aspide is far from being effective.
%
%%Moreover, several graphic environment (e.g., {\em APE}~\cite{Sureshkumar-etal-07}, 
%%{\em Digg}\footnote{\url{http://www.ezul.net/2010/09/gui-for-dlv.html},
%%{\em Visual DLV}~\cite{perri-etal-07-sea}, {\em DLV!sual}\footnote{\url{http://thp.io/2009/dlvisual},
%and {\em OntoDLV} ~\cite{ricc-etal-2008-jlc})

In this paper we present a pragmatic solution for testing ASP programs. %provides a contribution in this setting. 
In particular, we present a new language for specifying and running {\em unit tests} on ASP programs. 
%Unit testing require to test individual units of source code to verify whether they behave as intended. 
%A unit is the smallest testable part of a program. 
%We intend as unit of an ASP programs $P$ any subset of the rules of $P$
%corresponding to a splitting set~\cite{lifs-turn-94} 
%(actually we exploit a generalization  of the splitting theorem 
%by Lifschitz and Turner~\cite{lifs-turn-94}
%to the non-ground case~\cite{eite-etal-2010}).
%In this way, the behavior of units can be verified 
%both when they run isolated from the original program, as well as 
%when they are left immersed in (or in part of) the original program.
%%
The testing language presented in this paper is inspired by the JUnit~\cite{junit} framework:
the developer can specify the rules composing one or several units, specify one or more inputs
and assert a number of conditions on both expected outputs and the expected behavior of sub-programs. 
The obtained test case specification can be run by exploiting an ASP solver, and
the assertions are automatically verified by analyzing the output of the chosen ASP solver.
Note that test case specification is applicable independently of the used ASP solver.
The testing language was implemented in \aspide, which also provides
the user with some graphic tools that make the development of test cases simpler.
The testing tool described in this work extends significantly the one formerly available
in \aspide, by both extending the language by more expressive (non-ground) assertions and
the support of weak-constraints, and enriching its collection of user-friendly graphical tools 
(including program composition, debugging, profiling, database management,
solver execution configuration, and output-handling) 
with a graphical test suite management interface.

As far as related work is concerned, the task of testing ASP programs was 
approached for the first time, to the best of our knowledge, in \cite{janh-2010,janh-2011}
where the notion of structural testing for ground normal ASP programs is defined and 
methods for automatically generating tests is introduced. 
The results presented in~\cite{janh-2010,janh-2011} are, somehow, orthogonal to the contribution of
this paper. Indeed,
%which are based on several notions of coverage 
%(i.e., measures of the way a set of test inputs covers some structural properties of the program).
no language/implementation is proposed in~\cite{janh-2010,janh-2011} for specifying/automatically-running 
the produced test cases; whereas, the language presented in this paper can be used 
for encoding the output of a test case generator based on the methods proposed in~\cite{janh-2010}.
Finally, it is worth noting that, testing approaches developed for other logic languages, 
like prolog~\cite{jack-96,plunit,prodt}, cannot be straightforwardly ported to ASP because 
of the differences between the languages.

\medskip

The rest of this paper is organized as follows: in Section~\ref{sec:aspide} we
overview \aspide; in section~\ref{sec:language} we introduce a language for specifying unit tests
for ASP programs; in Section~\ref{sec:testgui} we describe the user interface components
of \aspide conceived for creating and running tests;
finally, in Section~\ref{sec:conclusion} we draw the conclusion.

%%%%%%%%%%%%%%%%%%%%%%%%%%%%%%%%%%%%%%%%%%%%%%%%%%%%%%%%%%%%%%%
\section{\aspide: Integrated Development Environment for ASP}\label{sec:aspide} 

\aspide is an Integrated Development Environment (IDE) for ASP, 
which features a rich {\em editing tool} with a collection of user-friendly 
{\em graphical tools} for ASP program development.
In this section we first summarize the main features of the system and 
then we overview the main components of the \aspide user interface.
For a more detailed description of \aspide, as well as for a complete comparison
with competing tools, we refer the reader to~\cite{febb-etal-lpnmr11}
and to the online manual published in the system web site 
\url{http://www.mat.unical.it/ricca/aspide}.

%%%%%%%
%\subsection{System Features}
\paragraph{System Features.}
\aspide is inspired by Eclipse, one of the most diffused programming environments.
The main features of \aspide are the following:
\begin{itemize}
\item {\em Workspace management}. 
The system allows one to organize ASP programs in projects, which are collected
in a special directory (called workspace).

\item {\em Advanced text editor}. The editing of ASP files is simplified by an advanced text editor.
Currently, the system is able to load and store ASP programs in the syntax of the 
ASP system DLV~\cite{leon-etal-2002-dlv}, and supports
the {\tt ASPCore} language profile employed in the ASP System Competition 2011 \cite{asp-comp-11}. \aspide can also manage {\em TYP files} specifying a mapping between program predicates and database tables 
in the \dlvdb syntax~\cite{terr-etal-2008}.
%The text editor of \aspide provides several functionalities.
Besides the core functionality that basic text editors offer
(like code line numbering, find/replace, undo/redo, copy/paste, etc.), 
\aspide offers other advanced functionalities, like:  
{\em Automatic completion},  {\em Dynamic code templates}, {\em Quick fix}, 
and {\em Refactoring}. 
Indeed, the system is able to complete (on request)
predicate names, as well as variable names. Predicate names are both learned while writing,
and extracted from the files belonging to the same project; variables
are suggested by taking into account the rule we are currently writing.
When several possible alternatives for completion are available the system shows a pop-up dialog.
Moreover, the writing of repeated programming patterns 
(like transitive closure or disjunctive rules for guessing the search space) 
is assisted by advanced auto-completion with code templates, which can 
generate several rules at once according to a known pattern.
Note that code templates can also be user defined by writing DLT~\cite{iann-etal-2003-asp} files. 
The refactoring tool allows one to modify in a guided way, among others, predicate names and variables (e.g., variable renaming in a rule is done by considering 
bindings of variables, so that variables/predicates/strings occurring 
in other expressions remain unchanged).
Reported errors or warnings can be automatically fixed by selecting (on request)
one of the system's suggested quick fixes, 
which automatically change the affected part of code.

\item {\em Outline navigation}. \aspide creates an outline view which graphically represents  
program elements.
Each item in the outline can be used to quickly access the corresponding line of code
(a very useful feature when dealing with long files),
and also provides a graphical support for building rules in the visual editor (see below).

\item {\em Dynamic code checking and error highlighting}.
Syntax errors and relevant conditions (like safety) are checked {\em while typing programs}: 
portions of code containing errors or warnings are immediately highlighted.
Note that the checker considers the entire project, and warns the user by indicating e.g., 
that atoms with the same predicate name have different arity in several files. This 
condition is usually revealed only when programs divided in multiple files are run together.

\item {\em Dependency graph}. 
The system is able to display several variants of the dependency graph associated 
to a program (e.g., depending on whether both positive and negative dependencies are considered).

\item {\em Debugger and Profiler}. Semantic error detection as well as code optimization  
can be done by exploiting graphic tools. 
In particular, we developed a graphical user interface for embedding 
in \aspide the debugging tool {\em spock}~\cite{brain-etal-07-sea}
(we have also adapted spock for dealing with the syntax of the \dlv system). 
Regarding the profiler, we have fully embedded the graphical interface presented 
in~\cite{cali-etal-2009}.

\item {\em Unit Testing}. The user can define unit tests and verify the behavior
of program units. The language for specifying unit tests,
as well as the graphical tools of \aspide assisting the development of tests,
are described in detail in the following sections.
% on the line with the principle of unit testing diffused in the development of
%software with iterative languages. 

\item {\em Configuration of the execution}. This feature allows one to configure and manage 
input programs and execution options (called {\em run configurations}).

\item {\em Presentation of results}. The output of the program (either answer sets, or 
query results) are visualized in a tabular representation or in a text-based console.
The result of the execution can be also saved in text files for subsequent analysis.

\item {\em Visual Editor}. The users can {\em draw} logic programs by exploiting 
a full graphical environment that offers a QBE-like tool 
for building logic rules~\cite{febbr-etal-10}.
The user can switch, every time he needs, from the text editor to the visual one (and vice-versa) 
thanks to a reverse-engineering mechanism from text to graphical format.

\item {\em Interaction with databases.}
Interaction with external databases is useful in several applications
(e.g.,~\cite{leon-etal-2005}).
\aspide provides a fully graphical import/export tool that automatically 
generates mappings by following the \dlvdb TYP file specifications~\cite{terr-etal-2008}.
Text editing of TYP mappings is also assisted by syntax coloring and auto-completion.
Database oriented applications can be run by setting \dlvdb as 
solver in a run configuration.
% A data integration scenario~\cite{leon-etal-2005}
%can be implemented by exploiting this feature.

\end{itemize}

\begin{figure}[t!]
\centering
\includegraphics[width=17.2cm,angle=90]{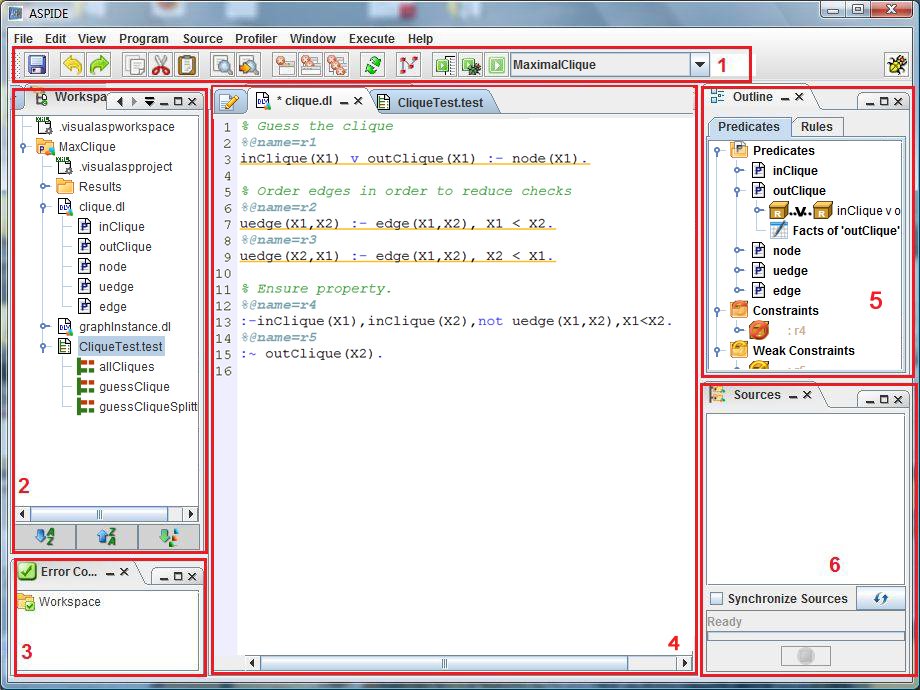} %,height=4.5cm
\caption{The \aspide graphical user interface.}\label{fig:mainWindow}
%\vspace{-0.5cm}
\end{figure}
%%%%%%%
%\subsection{Interface Overview}
\paragraph{Interface Overview}
The user interface of \aspide is depicted in Figure~\ref{fig:mainWindow}.
The most common operations can be quickly executed through a toolbar
present in the upper part of the GUI (zone 1).
From left to right there are buttons allowing to: save files, undo/redo, 
copy \& paste, find \& replace, switch between visual to text editor, 
run the solver/profiler/debugger.
The main editing area (zone 4) is organized in a multi-tabbed panel 
possibly collecting several open files. On the left there is the explorer panel (zone 2)
which allows one to browse the workspace; and the error console (zone 3).  The explorer panel lists projects and files included in the workspace,
while the error console organizes errors and warnings according to the project and files where they are localized.
On the right, there are the outline panel (zone 5) and the sources panel (zone 6).
The first shows an outline of the currently edited file, while the latter reports a list of the database sources connected with the current project.
Note that, the layout of the system can be customized by the user,
indeed panels can be moved and rearranged. % as the user likes.

\aspide is written in Java and runs on the most diffused operating systems
(Microsoft Windows, Linux, and Mac OS) and can connect to any database 
supporting Java DataBase Connectivity (JDBC).
%Currently, \aspide can deal with ASP programs in the syntax of the 
%ASP system DLV~\cite{leon-etal-2002-dlv}, and supports
%the {\tt ASPCore} language profile employed in the ASP System Competition 2011~\cite{asp-comp-11}.
%Data base management is compliant with \dlvdb~\cite{terr-etal-2008} language directives.

%%%%%%%%%%%%%%%%%%%%%%%%%%%%%%%%%%%%%%%%%%%%%%%%%%%%%%%%%%%%%%%%%%%%%
\section{A language for testing ASP programs}\label{sec:language}
Software testing~\cite{somm-book} is an activity aimed at evaluating the behavior of a program 
by verifying whether it produces the required output for a particular input.
The goal of testing is not to provide means for establishing 
whether the program is totally correct; conversely testing is a pragmatic 
and cheap way of finding errors by executing some test.
A test case is the specification of some input $I$ and corresponding expected outputs $O$.
A test case fails when the outputs produced by running the program does not correspond to $O$, it passes otherwise.

One of the most diffused white-box%
%%%
\footnote{A test conceived for verifying some functionality of an application without knowing the code
internals is said to be a black-box test. A test conceived for verifying the behavior of a
specific part of a program is called white-box test. White box testing is an activity usually
carried out by developers and is a key component of agile software development~\cite{somm-book}.}
%%%
testing techniques is {\em unit testing}. 
%
%used in the design of tests: a low detail level
%allows to define assertions on a software without any knowledge of internal 
%implementation then the programmer can tests whole program (black-box testing);
%increasing detail level, the programmer can define tests on 
%parts of program called {\em units}.
The idea of unit testing is to assess an entire software by testing its subparts
called {\em units} (and corresponding to small testable parts of a program).
%, and 
%unit testing require to test individual units of source code 
%to verify whether they behave as intended. 
In a software implemented by using imperative object-oriented languages, 
unit testing corresponds to assessing separately portions of the code like 
class methods.
The same idea can be applied to ASP, once the notion of unit is given.
We intend as unit of an ASP programs $P$ any subset of the rules of $P$
corresponding to a splitting set~\cite{lifs-turn-94} 
(actually the system exploits a generalization of the splitting theorem 
by Lifschitz and Turner~\cite{lifs-turn-94}
to the non-ground case~\cite{eite-etal-2010}).
In this way, the behavior of units can be verified 
(by avoiding unwanted behavioral changes due to cycles)
both when they run isolated from the original program as well as 
when they are left immersed in (part of) the original program.

In the following, we present a pragmatic solution for testing ASP programs,
which is a new language, inspired by the JUnit~\cite{junit} framework, 
for specifying and running {\em unit tests}. % on ASP programs. 
The developer, given an ASP program, can select the rules composing a unit, 
specify one or more inputs, and assert a number of conditions on the expected output. 
The obtained test case specification can be run, and
the assertions automatically verified by calling an ASP solver and checking its output.
In particular, we allow three test execution modes:
\begin{itemize}
\item {\em Execution of selected rules}. The selected rules will be executed 
separated from the original program on the specified inputs.
\item {\em Execution of split program}. The program corresponding to the 
splitting set containing the atoms of the selected rules is run and tested. 
In this way, the "interface" between two 
splitting sets can be tested (e.g., one can assert some expected properties
on the candidates produced by the guessing part of a program 
by excluding the effect of some constraints in the checking part).
%On the selected rules that respect the properties
%of the previous point, the system includes also the part of program where, for each rule,
%at least one atom of the head can be unified with some atom of the splitting set and no
%atoms of the body can be unified with any atom of the splitting set (unification of atoms
%makes sense on splitting of non-ground programs, see~\cite{eite-etal-2010}). Making this union
%of rules the meaning is that we want to execute the slice giving, as input, the part of
%program that we are including. The same result is obtained if we execute alone the included
%rules and give the results as input to the slice (see ~\cite{lifs-turn-94} for an
%exhaustive explanation of the \emph{splitting sequence theorem}).
\item {\em Execution in the whole program}. The original program is run 
and specific assertions regarding predicates contained in the unit are checked.
This corresponds to filtering test results on the atoms contained in the selected rules.
\end{itemize}

%%%%%%%%%%%%%%%%%%%%%%
\paragraph{Testing Language.}
A test file can be written according to the following grammar:%
\footnote{Non-terminals are in bold face; token specifications are omitted for simplicity.}
\\
%\begin{footnotesize}
{\scriptsize

\hsp {\tt 1 :  invocation("{\bf invocationName}" [ ,"{\bf solverPath}", "{\bf options}" ]?);}

\hsp {\tt 2 :  [ [ input("{\bf program}"); ] | [ inputFile("{\bf file}"); ] ]* }

\hsp {\tt 3 :  [}

\hsp {\tt 4 :  {\bf testCaseName}([ SELECTED\_RULES | SPLIT\_PROGRAM | PROGRAM ]?) }

\hsp {\tt 5 :  \{}

\hsp {\tt 6 :    [newOptions("{\bf options}");]?}

\hsp {\tt 7 :    [ [ input("{\bf program}"); ] | [ inputFile("{\bf file}"); ] ]* }

\hsp {\tt 8 :   [ [ excludeInput("{\bf program}"); ] 

\hsp {\tt 9 :   | [ excludeInputFile("{\bf file}"); ] ]* }

\hsp {\tt 10 :    [ }

\hsp {\tt 11 :  [ filter | pfilter | nfilter ] }
                         
\hsp {\tt 12 :         [ [ ({\bf predicateName} [ ,{\bf predicateName} ]* ) ] }

\hsp {\tt 13 :         | [SELECTED\_RULES] ] ;}

\hsp {\tt 14 :  ]?}

\hsp {\tt 15 :  [ selectRule({\bf ruleName}); ]*}

\hsp {\tt 16 :  [ [ {\bf assertName}( [ {\bf intnumber}, ]? [ [ "{\bf atoms}" ] | [ "{\bf constraint}" ] ); ]}

\hsp {\tt 17 :  | [ assertBestModelCost({\bf intcost} [, {\bf intlevel} ]?  ); ] ]* }

\hsp {\tt 18 : \} }

\hsp {\tt 19 : ]* }

\hsp {\tt 20 : [ [ {\bf assertName}( [ {\bf intnumber}, ]? [ [ "{\bf atoms}" ] | [ "{\bf constraint}" ] ); ]}

\hsp {\tt 21 : | [ assertBestModelCost({\bf intcost} [, {\bf intlevel} ]?  ); ] ]* }\\ \
}}

A test file might contain a single test or a test suite (a set of tests) including several test cases. 
Each test case includes one or more assertions on the execution results.

The \emph{invocation} statement (line 1) sets the global invocation settings, that
apply to all tests specified in the same file (name, solver, and execution options).
In the implementation, the invocation name might correspond to an \aspide run configuration,
and the solver path and options are not mandatory.

The user can specify one or more global inputs by writing some 
\emph{input} and \emph{inputFile} statements (line 2).
The first kind of statement allows one for writing  the input
of the test in the form of ASP rules or simply facts; 
the second statement indicates a file that contains some input in ASP format.

A test case declaration (line 4) is composed by a name and an optional parameter that allows one to
choose if the execution will be done on the entire program, on a subset of rules, or 
on the program corresponding to the splitting set containing the selected rules.
The user can specify particular solver options (line 6), as well as certain inputs (line 7)
which are valid in a given test case. Moreover, global inputs of the test suite 
can be excluded by exploiting \emph{excludeInput} and \emph{excludeInputFile} statements
(lines 8 and 9). 
The optional statements \emph{filter}, \emph{pfilter} and \emph{nfilter} (lines 11, 12, and 13) are used to 
filter out output predicates from the test results 
(i.e., specified predicate names are filtered out from the results when the assertion is executed).%
\footnote{\emph{pfilter} selects only positive literals and excludes the strongly negated ones, while 
\emph{nfilter} has opposite behavior.}
The statement \emph{selectRule} (line 15) allows one for selecting rules among the ones composing 
the global input program.
A rule $r$ to be selected must be identified by a name, which is expected to be specified in the input program
in a comment appearing in the row immediately preceding $r$ (see Figure~\ref{fig:mainWindow}). 
\aspide adds automatically the comments specifying rule names.
If a set of selected rules does not belong to the same splitting set, the system has to print a warning
indicating the problem.

The expected output of a test case is expressed in terms of assertion statements (lines 16/21).
The possible assertions are:

\begin{itemize}
\item {\em assertTrue(''atomList'')}/{\em assertCautiouslyTrue(''atomList'')}. Asserts that all atoms of the atom list must be true in any answer sets;
\item {\em assertBravelyTrue(''atomList'')}. Asserts that all atoms of the atom list must 
be true in at least one answer set;
\item {\em assertTrueIn(number, ''atomList'')}. Asserts that all atoms of 
the atom list must be true in exactly \emph{number} answer sets;
\item {\em assertTrueInAtLeast(number, ''atomList'')}. Asserts that all atoms 
of the atom list must be true in at least \emph{number} answer sets;
\item {\em assertTrueInAtMost(number, ''atomList'')}. Asserts that all atoms of 
the atom list must be true in at most \emph{number} answer sets;
\item {\em assertConstraint('':-constraint.'')}. Asserts that all answer sets 
must satisfy the specified constraint;
\item {\em assertConstraintIn(number, '':-constraint.'')}. Asserts that exactly \emph{number} answer sets
must satisfy the specified constraint;
\item {\em assertConstraintInAtLeast(number, '':-constraint.'')}. Asserts that at least \emph{number} answer sets
must satisfy the specified constraint;
\item {\em assertConstraintInAtMost(number, '':-constraint.'')}. Asserts that at most \emph{number} answer sets
must satisfy the specified constraint;
\item {\em assertBestModelCost(intcost)} and {\em assertBestModelCost(intcost, intlevel)}. In case of execution of
programs with weak constraints, they assert that the cost of the best model with level \emph{intlevel} must be \emph{intcost};
\end{itemize}

\noindent together with the corresponding negative assertions: \emph{assertFalse}, 
\emph{assertCautiouslyFalse}, \emph{assertBravelyFalse}, \emph{assertFalseIn}, 
\emph{assertFalseInAtLeast}, \emph{assertFalseInAtMost}.
The \emph{atomList} specifies a list of atoms that can be ground or non-ground; in
the case of non-ground atoms the assertion is true if some ground instance matches in some/all answer sets. 
Assertions can be global (line 20-21) or local to a single test (line 16-17).

In the following we report an example of test case. 
%a test suite in order to test the Hamintonian Path problem.

\paragraph{Test case example.}
The maximum clique is a classical hard problem in graph theory requiring 
to find the largest  clique (i.e., a complete subgraph of maximal size)  
in an undirected graph.
Suppose that the graph {\em G} is specified by using facts over predicates 
{\em node} (unary) and {\em edge} (binary), then
the program in Figure~\ref{fig:mainWindow} solves the problem.\\ \

%\newcommand{\hsp }{\noindent \hspace{0.5cm}}
%{
%\footnotesize
%\hsp {\tt \% Guess the clique}
%
%\hsp {\tt r$_{1}$: inClique(X1) v outClique(X1) :- node(X1).}
%
%\hsp {\tt \% Order edges in order to reduce checks}
%
%\hsp {\tt r$_{2}$: uedge(X1,X2) :- edge(X1,X2), X1 < X2.}
%
%\hsp {\tt r$_{3}$: uedge(X2,X1) :- edge(X1,X2), X2 < X1.}
%
%\hsp {\tt \% Ensure property. }
%
%\hsp {\tt  r$_{4}$: :- inClique(X1), outClique(X2), not uedge(X1,X2), X1 < X2.}
%
%\hsp {\tt  r$_{5}$: :\~\  outClique(X2).} \\ \ 
%}

\begin{figure}[t!]
\centering
\includegraphics[width=6cm]{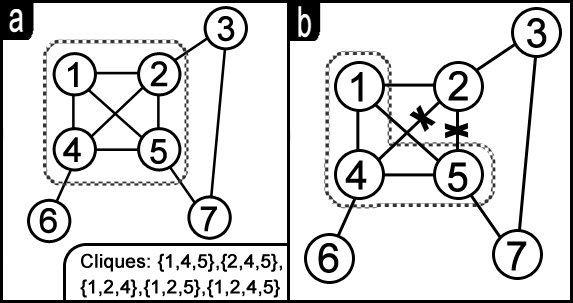}
\caption{Input graphs.}\label{fig:graphExample}
\end{figure}

\noindent The disjunctive rule ({\tt r$_{1}$}) guesses a subset $S$ of the nodes to be
in the clique, while the rest of the program checks whether $S$
constitutes a clique, and the weak constraint ({\tt r$_{5}$}) maximizes the size of $S$. 
Here, an auxiliary predicate {\em uedge}  exploits an ordering for reducing the time spent in checking.
%
%We now define a test suite file, based on a simple instance.
%
Suppose that the encoding is stored in a file named \emph{clique.dl}; and 
suppose also that the graph instance, composed by facts 
\{ \emph{node(1). node(2). node(3). node(4). node(5). node(6). node(7). 
edge(1,2). edge(2,3). edge(2,4). edge(1,4). edge(1,5). edge(4,5). 
edge(2,5). edge(4,6). edge(5,7). edge(3,7).}\},
is stored in the file named \emph{graphInstance.dl}
(the corresponding graph is depicted in Figure~\ref{fig:graphExample}a).
The following is a simple test suite specification for the above-reported ASP program:\\ \

{
\scriptsize
\hsp {\tt invocation("MaximalClique", "/usr/bin/dlv", "");}

\hsp {\tt inputFile("clique.dl");}

\hsp {\tt   inputFile("graphInstance.dl");}

\hsp {\tt maximalClique()}

\hsp {\tt \{}

\hsp {\tt   assertBestModelCost(3);}

\hsp {\tt \}}

\hsp {\tt constraintsOnCliques()}

\hsp {\tt \{}

\hsp {\tt excludeInput(":\~\ outClique(X2).");}

\hsp {\tt assertConstraintInAtLeast(1,":- not inClique(1), not inClique(4).");}

\hsp {\tt assertConstraintIn(5,":- \#count\{ X1: inClique(X1) \} < 3.");}

\hsp {\tt \}}

\hsp {\tt checkNodeOrdering(SELECTED\_RULES)}

\hsp {\tt \{}

\hsp {\tt   inputFile("graphInstance.dl");}

\hsp {\tt   selectRule("r2");}

\hsp {\tt   selectRule("r3");}

\hsp {\tt   assertFalse("uedge(2,1).");}

\hsp {\tt \}}

\hsp {\tt guessClique(SPLIT\_PROGRAM)}

\hsp {\tt \{}

\hsp {\tt  selectRule("r1");}

\hsp {\tt  assertFalseInAtMost(1,"inClique(X).");}

\hsp {\tt  assertBravelyTrue("inClique(X).");}

\hsp {\tt \}}

}

\begin{figure}[t!]
\centering
\includegraphics[width=\textwidth]{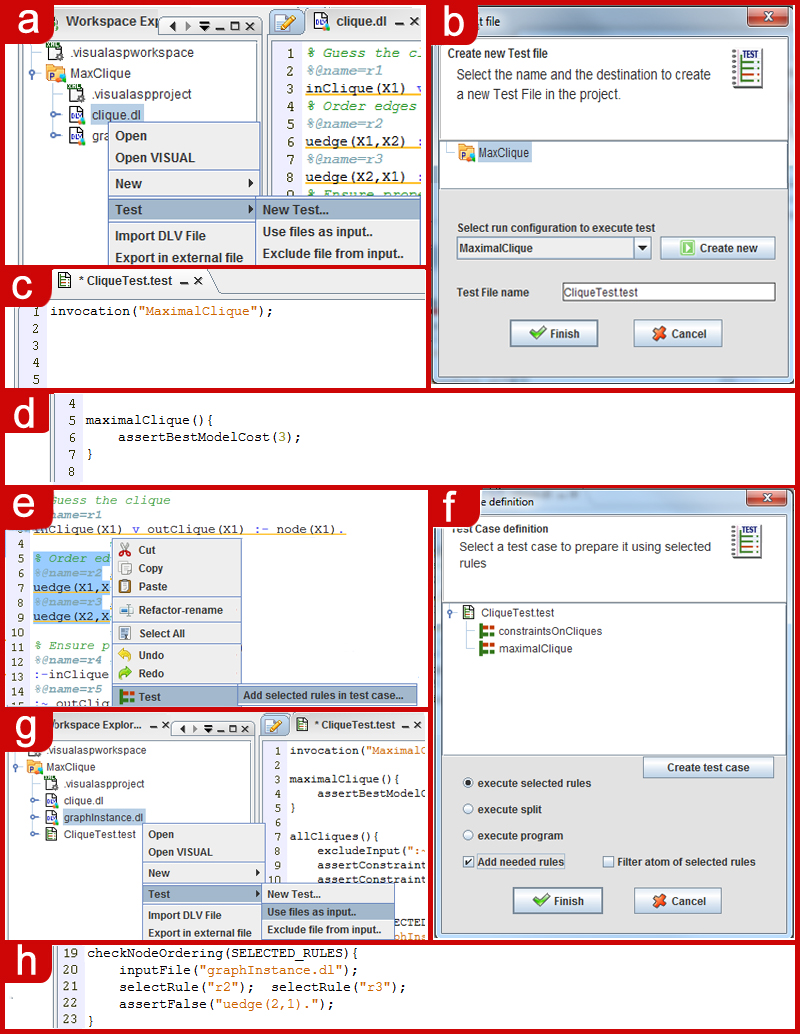}
\caption{Test case creation.}\vspace{-0.2cm}
\label{fig:screenSea1}
\end{figure}

Here, we first set the invocation parameters by indicating \dlv as solver, then we specify
the file to be tested \emph{clique.dl} and the input file \emph{graphInstance.dl},
by exploiting a global input statement;
then, we add the test case \emph{maximalClique}, in which we assert that
the best model is expected to have a cost (i.e., the number of weak constraint violations
corresponding to the vertexes out of the clique) of 3 (\emph{assertBestModelCost(3)} in Figure~\ref{fig:screenSea1}).

In the second test case, named \emph{constraintsOnCliques}, we require that $(i)$ vertexes 1 and 4 belong 
to at least one clique, and $(ii)$ for exactly five answer sets the size of the corresponding clique 
is greater than 2. (The weak constraint is removed to ensure the computation of all cliques by \dlv.)

In the third test case, named \emph{checkNodeOrdering}, 
we select rules \emph{r$_{2}$} and \emph{r$_{3}$}, and 
we require to test selected rules in isolation, discarding all the other statements of the input.
We are still interested in considering ground facts 
that are included locally (i.e., we include the file \emph{graphInstance.dl}).
In this case we assert that \emph{uedge(2,1)} is false, 
since edges should be ordered by rules \emph{r$_{2}$} and \emph{r$_{3}$}.

Test case \emph{guessClique} is run in \emph{SPLIT\_PROGRAM} 
modality, which requires to test the subprogram containing all the rules belonging to the splitting set 
corresponding to the selection (i.e., \emph{\{inClique, outClique, node\}}).
In this test case the sub-program that we are testing is composed by the disjunctive rule 
and by the facts of predicate \emph{node} only. Here we require that
there is at most one answer set modeling the empty clique, and 
there is at least one answer set modeling a non-empty clique.

The test file described above can be created graphically and executed in \aspide as described
in the following section.

\begin{figure}[t!]
\centering
\includegraphics[width=\textwidth]{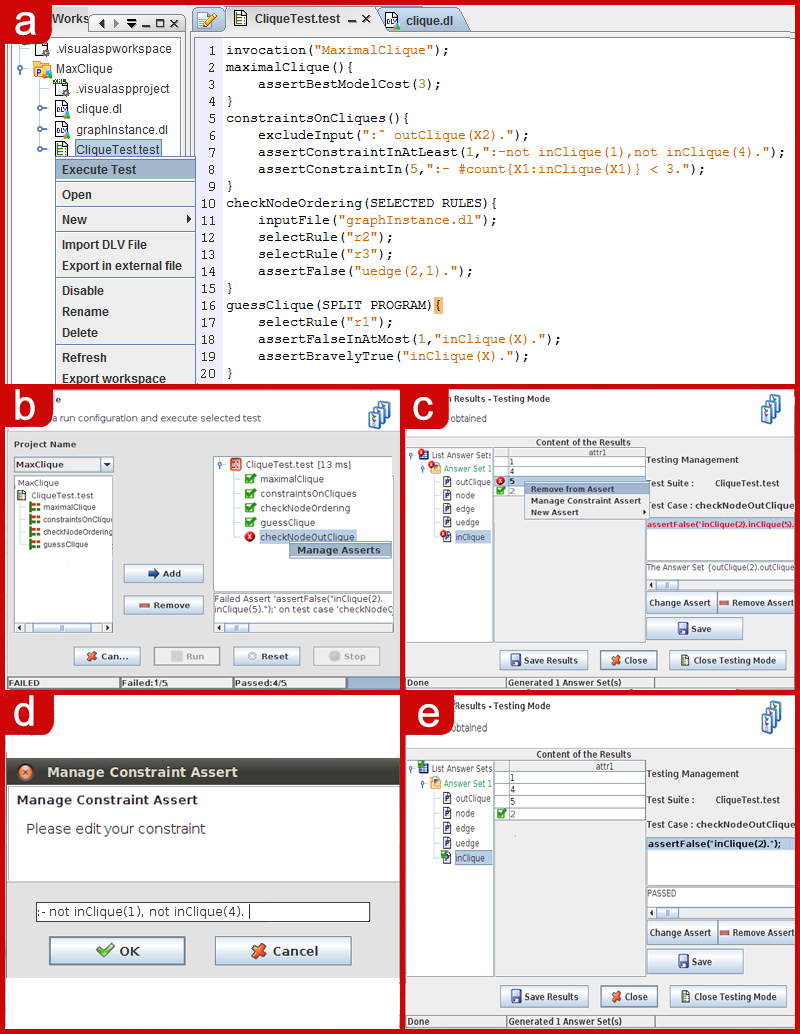}
\caption{Test case execution and assertion management.}
\vspace{-0.2cm}
\label{fig:screenSea2}
\end{figure}

%%%%%%%%%%%%%%%%%%%%%%
\section{Unit Testing in \aspide}\label{sec:testgui}

In this section we describe the graphic tools implemented in \aspide conceived 
for developing and running test cases. Space constraints prevent us from providing
a complete description of all the usage scenarios and available commands.
However, in order to have an idea about the capabilities of the testing interface of \aspide,
we describe step by step how to implement the example illustrated in the previous section. 

Suppose that we have created in \aspide a project named MaxClique, 
which contains the files \emph{clique.dl} and \emph{graphInstance.dl}
(see Fig.~\ref{fig:mainWindow}) storing 
the encoding of the maximal clique problem and the graph instance presented in the previous section, respectively. 
Moreover we assume that both input files are included in a run configuration named {\em MaximalClique},
and we assume that the \dlv system is the solver of choice in {\em MaximalClique}.
Since the file that we want to test in our example is \emph{clique.dl}, 
we select it in the \emph{workspace explorer}, then we click the right button of 
the mouse and select \emph{New Test} from the popup menu (Fig.~\ref{fig:screenSea1}a).
The system shows the test creation dialog (Fig.~\ref{fig:screenSea1}b), 
which allows one for both setting the name of the test file and selecting a previously-defined 
run configuration (storing execution options and input files). 
By clicking on the \emph{Finish} button, the new test file is created (see Fig.~\ref{fig:screenSea1}c)
where a statement regarding the selected run configuration is added automatically.
We add the first unit test (called {\em maximalClique}) by exploiting the text editor
(see Fig.~\ref{fig:screenSea1}d), whereas we build the remaining ones 
(working on some selected rules) by exploiting the logic program editor.
After opening the  \emph{clique.dl} file, we select rules $r_{2}$ and $r_{3}$ 
inside the text editor, we right-click on them and we select \emph{Add selected rules in test case} 
from the menu item \emph{Test} of the popup menu (fig.~\ref{fig:screenSea1}e). 
The system opens a dialog window where we indicate the test file in which we want to add 
the new test case (fig.~\ref{fig:screenSea1}f). We click on the \emph{Create test case};
the system will ask for the name of the new test case and we write \emph{guessClique}; after that, on the window, 
we select the option \emph{execute selected rules} and click on the \emph{Finish} 
button. The system will add the test case \emph{guessClique} filled with the 
\emph{selectRule} statements indicating the selected rules. 
To add project files as input of the test case, we select them from the 
\emph{workspace explorer} and click on \emph{Use file as input} 
in the menu item \emph{Test} (fig.~\ref{fig:screenSea1}g).
We complete the test case specification by adding the assertion, thus
the test created up to now is shown in figure~\ref{fig:screenSea1}h.
Following an analogous procedure we create the remaining test cases (see Fig.~\ref{fig:screenSea2}a). 
% \emph{guessSliceNonReachability}
%and the final global assert \emph{assertFalse("inPath(1, 2).");}.
%
To execute our tests, we right-click on the test file and select \emph{Execute Test}. 
The \emph{Test Execution Dialog} appears and the results are shown to the programmer (see Fig.~\ref{fig:screenSea2}b). Failing tests are indicated by a red icon, while green icons 
indicate passing tests. At this point we add the following additional test: \\

{\scriptsize
\hsp {\tt checkNodeOutClique() \ }

\hsp {\tt  \{ }

\hsp {\tt  excludeInput("edge(2,4).edge(2,5)."); }

\hsp {\tt  assertFalse("inClique(2). inClique(5)."); }

\hsp {\tt  \} }
}\\
%The meaning of this test case is the following: looking the graph of the 
%figure~\ref{fig:graphExample}c we choose to start from node 1 for the Hamiltonian Path and 
%to remove the edge from node 4 to node 5, so the node 5 will not be reachable. In the 
%encoding of the problem there is the constraint \emph{":- vtx(X), not reached(X).} ensuring 
%that all nodes must be reached; however with this constraint, no solutions will be generated 
%so we decide to exclude this constraint on the execution of the test case to generate some 
%solutions (that will not be Hamiltonian path solutions). We know that the node 5 will not 
%be reached, but suppose the user wrongs the assertion and asserts that all the Answer Sets 
%must contain \emph{reached(5)}, the test case will fail and the \emph{Test Execution Dialog}  

This additional test (purposely) fails, this can be easily seen by looking at
Figure~\ref{fig:graphExample}b; and the reason for this failure is indicated (see Fig. ~\ref{fig:screenSea2}b)
in the test execution dialog.
In order to know which literals of the solution do not satisfy the assertion, 
we right-click on the failed test and select \emph{Manage Asserts} from the menu.
A dialog showing the outputs of the test appears where, in particular, predicates and literals 
matching correctly the assertions are marked in green, whereas the ones 
violating the assertion are marked in red 
(gray icons may appear to indicate missing literals which are expected to be in the solution).
In our example, the assertion is \emph{assertFalse(''inClique(2). inClique(5).'')}; 
however, in our instance, node 5 is contained in the maximal clique composed by nodes \emph{1, 4, 5}; 
this is the reason for the failing test. Assertions can be modified graphically, and, in this case, we act  directly 
on the result window (fig.~\ref{fig:screenSea2}c). 
We remove the node 5 from the assertion by selecting it; moreover we right-click on the instance of \emph{inClique}
that specifies the node 5 and we select \emph{Remove from Assert}. The atom \emph{node(5)} will be removed
from the assertion and the window will be refreshed showing that
the test is correctly executed (see fig.~\ref{fig:screenSea2}e).
The same window can be used to manage constraint assertions; in particular, by clicking on 
\emph{Manage Constraint Assert} of the popup menu, a window appears that allows the user to set/edit constraints 
(see fig.~\ref{fig:screenSea2}d).

%%%%%%%%%%%%%%%%%%%%%%%%%%%%%%%%%%%%%%%%%%%%%%%%%%%%%%%%%%%%%%%
\section{Conclusion}\label{sec:conclusion}

%In this paper we described the new testing environment of \aspide.
This paper presents a pragmatic environment for testing ASP programs. %provides a contribution in this setting. 
In particular, we propose a new language, inspired by the JUnit~\cite{junit} framework, 
for specifying and running {\em unit tests} on ASP programs.
The testing language is general and suits both the DLV~\cite{leon-etal-2002-dlv} and clasp~\cite{gebs-etal-2007-ijcai} ASP dialects.
The testing language has been implemented in \aspide together with some 
graphic tools for easing both the development of tests and the analysis of test execution (via DLV).

%In particular, we introduced a new language for specifying unit tests on ASP programs,
%
%
%The testing language is  and allows developers 
%to specify and test individual units of ASP code composing a program, 
%with the aim of determining if they work as expected.
%The developer can assert a number of conditions on the expected output of 
%program units. The behavior of program units can be verified 
%both when they run isolated from the original program, as well as 
%when they are left immersed in the original program.

%The new testing environment enriches the collection of user-friendly graphical tools 
%available in \aspide for program composition, debugging, profiling, database management,
%solver execution configuration, and output-handling.

%\medskip

As far as future work is concerned, we plan to extend \aspide 
by improving/intro\-ducing additional dynamic editing instruments, and graphic tools like
VIDEAS ~\cite{oetsch-etal-2011}.
Moreover, we plan to further improve the testing tool by supporting (semi)automatic 
test case generation based on the structural testing techniques proposed in~\cite{janh-2010,janh-2011}.

\medskip
{\noindent {\bf Acknowledgments.} This work has been partially supported by the Calabrian Region
under PIA (Pacchetti Integrati di Agevolazione industria, artigianato e servizi) project DLVSYSTEM approved in BURC n. 20  parte III del 15/05/2009 - DR n. 7373 del 06/05/2009.}

\bibliographystyle{splncs}
%\bibliography{bibtex}

\newcommand{\SortNoOp}[1]{}

\end{document}